\documentclass[letterpaper, 10 pt, conference]{ieeeconf}  

\IEEEoverridecommandlockouts                              

\usepackage{graphics} 
\usepackage{epsfig} 
\usepackage{mathptmx} 
\usepackage{times} 
\usepackage{amsmath} 
\usepackage{amssymb}  

\newcommand{\eg}{e.\,g.,\ }
\newcommand{\ie}{i.\,e.,\ }

\newcommand{\etc}{{etc}.\ }

\title{\LARGE \bf
Improving Take-over Situation by Active Communication
}

\author{Monika Sester$^{1}$, 
Mark Vollrath$^{2}$,
Hao Cheng$^{1}$
\thanks{$^{1}$Monika Sester and Hao Cheng are with the Institute of Cartography and Geoinformatics, Leibniz University Hannover, Germany,
        {\tt\small Monika.Sester@uni-hannover.de; Hao.Cheng@uni-hannover.de}}%
\thanks{$^{2}$Mark Vollrath is with the Institute of Psychology, Technische Universit{\"a}t Braunschweig, Gaussstrasse 23, 38106 Braunschweig, Germany
        {\tt\small Mark.Vollrath@tu-bs.de}}%
}

\begin{document}

\maketitle
\thispagestyle{empty}
\pagestyle{empty}

\begin{abstract}

In this short paper an idea is sketched, how to support drivers of an autonomous vehicle in taking back control of the vehicle after a longer section of autonomous cruising. The hypothesis is that a clear communication about the location and behavior of relevant objects in the environment will help the driver to quickly grasp the situational context and thus support drivers in safely handling the ongoing driving situation manually after take-over. Based on this hypothesis, a research concept is sketched, which entails the necessary components as well as the disciplines involved.  

\end{abstract}

\section{INTRODUCTION}
\label{sec:introduction}

In a few years, SAE Level 3~\cite{SAE_j3016b_2018} automatic vehicles will be on the market. 
They will perform certain driving tasks, but the driver has to take over control and resume manual driving whenever the system requires her/him to do so. This is especially the case in situations where the automatic vehicle reaches a system limit, for example, due to bad weather conditions impairing the sensor performance, complex situations like traffic jams, construction works \etc (see \cite{klamroth2019transitionen}).
Thus, a major problem with the introduction of autonomous vehicles (AV)s is to organize the handover between autonomous and driver-controlled driving. As described, supporting the driver in these situations is required as these take-over situations will most likely be complex, demanding situations~\cite{klamroth2019transitionen}.
If no such support is given, it might pose a stress for the user knowing that at any instance, she/he would be asked to take over again. This might lead to a feeling of insecurity and uneasiness, leading to a potential non-use of autonomous facilities. 
In addition, the driver may not trust the autonomous facilities when such take-overs happen unexpectedly.

There is large ongoing discussion about the appropriate duration for taking over control.
Earlier studies recommended 3 to 5 seconds, based on the mean duration of take-over times of expert drivers just waiting to take over. 
However, it has to be ensured that every kind of driver (aged, distracted, fatigued, etc.) has to have sufficient time for take-over.
Additionally, when looking at gaze behavior there is some indication that even when drivers had taken over control, they need additional time until they have really perceived all relevant objects in this situation
~\cite{klamroth2019transitionen}.
This was also supported by in-depth analyses of take-over strategies which demonstrate that driver requires additional time to really achieve an adequate level of situation awareness~\cite{varytimidis2018action}.

Situation awareness describes the fundamental stages of obtaining an adequate understand of the current situation required to safely handle this situation~\cite{endsley1988design}. 
It includes the perception of all relevant elements (stage 1), the understanding of their meaning (stage 2) and the ability to anticipate what will happen in the next few seconds (stage 3). 
Thus, when taking over control from an automatic vehicle the driver has to scan the environment for relevant elements (mainly other cars or trucks, but also traffic signs, the position of the car on the road, construction work, obstacles, weather conditions which require an adaptation of speed etc.) and understand whether these are relevant for their driving task in the near future. 
Especially after having been distracted before the take-over (\eg working, typing, watching videos) this is not an easy task within the typical time budget given to take-over control. 
Supporting the driver in this orientation tasks is thus highly relevant for traffic safety, but also the confidence and trust in automatic cars and the resulting usage of these features. As the concept of situation awareness points out, this comprises to perceive all relevant elements of the current driving situation and understand their meaning, \ie their relevance to the driving task after take-over when driving manually again. 
The next section describes the approach developed to achieve this aim.

\section{Approach}
\label{sec:approach}

When the driver has to take control of the vehicle back, she/he has to be aware of the current traffic situation. 
In order to get this overview, this is typically done by looking around, \ie looking in the mirror, over the shoulder. 
The idea in this concept paper is to mimic this process and assist users by providing them with additional knowledge about objects in their immediate environment and their potential behavior. 
In this way, also information about occluded object, objects, which have not been noticed, or objects, which potentially represent a danger will be visualized or communicated.

The first step in this support is to automatically determine the relevant elements of the traffic situation. 
The second step is to communicate them to the driver in an efficient, easy to understand way while not only supporting the perception of these elements, but also their meaning. 
With regard to that, the driver has to decide whether a certain element requires an action or adaptation of the driving task. 
For example, detecting an obstacle 100\,m ahead in the lane of the driver and dense traffic on the adjacent lanes would require the driver to slow down and come to a stand-still. 
Thus, the presentation of the relevant elements should be done in a way which makes it obvious for the driver what to do by, for example, indicating adequate actions required.

Thus, the system acts as a kind of omniscient, friendly co-driver, helping the driver to quickly master the situation, by directing the gaze and attention to the essential elements and by providing action cues which enable a safe manual handling of this situation. 
Those elements are, \eg 

\begin{itemize}
    \item static scene / road configuration (traffic regulations, junctions in the vicinity, etc.)
    \item mobile objects and their past and predicted movements
    \item dangerous situations in the vicinity (\eg road slippage, faded street markings, accidents)
\end{itemize}

This analysis will exploit the environmental sensors of the vehicle, in conjunction with map information. 
Whereas the static information can be gained from digital maps, the dynamics (as well as the updated information) will be determined from the vehicle sensors (LiDAR, images, etc.), potentially also through car2X-communication from other sensors. 

\subsection{Research hypothesis and research challenges}
\label{sucsec:reseach_hypothesis_questions}

The research hypothesis is: 

It is possible to improve the quality of take-over of control after a longer ride in an autonomous system and to support manual driving directly after the take-over, thus resulting in a higher level of trust in the automation, a larger acceptance of these systems and also a better traffic safety, by providing the driver with exactly the information needed to take over and drive safely manually after the take-over. 

This leads to several challenges, which have to be addressed:
\begin{itemize}
    \item Which objects of the environment (static/dynamic) and what component of their behavior are relevant for an activity of a driver at a certain moment in a certain situation?
    \item Which elements require an adaptation of the driving performance?
    \item How can the attention of the driver be guided to these relevant elements and how can the understanding of their meaning be supported, including indications of what the driver should do when driving manually?
    \item Is it sufficient to just communicate/present the above described elements independently, or do they have to be combined to a process, or sequence of actions? (``first check driver in front of you, then have an eye on the car approaching from behind with high velocity")?
\end{itemize}
    
\section{CHALLENGES}
\label{sec:challenges}
\begin{figure}[tb]
\centering
\includegraphics[trim=0.1in 0.2in 0.3in 0.1in, clip=true, width=1\linewidth]{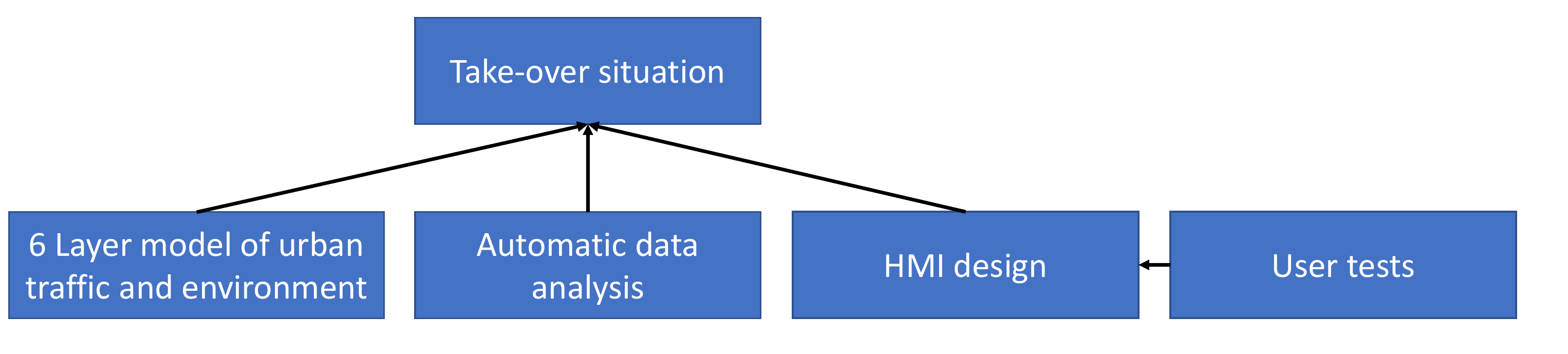}
\caption{The conceptual structure of take-over situation improvement.}
 \label{fig:pipeline}
\end{figure}

These questions have to be answered in collaboration among engineers, psychologists and computer scientists. 
Fig.~\ref{fig:pipeline} provides an overview of the necessary components that are key to the improvement of take-over situation.

\subsection{Selection of relevant elements of traffic situation}
A selection of information elements has to be made, which is needed to adequately react in a traffic situation. A good starting point is the 6 layer model of urban traffic and environment~\cite{scholtes20216}. Namely, L1 - road network and traffic guidance objects, L2 - roadside structures, L3 - temporary modifications of L1 and L2, L4 - dynamic objects, L5 - environmental conditions, and L6 - digital information.
This 6 layer model has to be adapted to the problem of immediate information need in a take-over situation.
Based on the resulting taxonomy of necessary elements, the next steps can be taken.
One example would be that, based on the information in the lower layers,  the top layer L6 can present prior information to a driver that a take-over is recommended when the situation does no longer allow for autonomous drive. 

\subsection{Automatic Data Analysis}
From the environmental sensors, objects can be extracted automatically, which are in the visible range of the sensors. 
To this end, modern Deep Learning approaches are very suitable, that can identify different, relevant object classes. 
The most important are other road users, such as cars and trucks, but also vulnerable road users like cyclists or pedestrians~\cite{Yurtsever2020Autonomous}. 
Moreover, with the help of object detection and gaze tracking, it is also possible to analyze whether the object is in the region of the driver's visual attention~\cite{jiang2018inferring}. Through car2car communication, also the information from other vehicles can be cooperatively exploited \cite{chen2019f}. 
Besides the recognition of static objects and dynamic objects including their trajectory, there are also approaches, which infer the intention and projected behavior of road users (\eg~\cite{alahi2016social,lee2017desire,gupta2018social,cui2019multimodal,park2020diverse,cheng2021amenet}). 
This can also help the driver estimate the traffic situation and understand the necessity of the take-over.

\subsection{HMI Design}
The presentation of the spatial information and the required actions can be conducted in several ways, including the visual and auditory modality, displays in the car, in the head-up display, but also AR visualization. This HMI design will be based on the theory of ecological interfaces which present information in a manner which directly enables correct actions of the driver (\eg \cite{schewe2021ecological}). 

\subsection{User tests}
For the HMI design, user tests will be conducted in a driving simulator. A set of typical take-over situations including possible events directly after the take-over has already been implemented in this situation for earlier studies (\eg \cite{vogelpohl2018transitioning,vogelpohl2020task}). 
Driving simulators enables a flexible design and presentation of possible HMI elements to support the driver in the manner presented above. In a first step, different concepts have to be developed with expert users and then implemented in these scenarios. User studies can then explore and demonstrate the efficiency of different HMI designs by comparing gaze behavior and driving performance directly after the take-over to a group of drivers taking over without any HMI support.

\section{Final Remarks and Outlook}
The acceptance of autonomous systems will largely depend on the ability of people to interact with them in a confident and appropriate manner. An essential element is a smooth hand-over between driving responsibilities. This paper addressed the challenges of this problem and proposes a way ahead towards a peaceful cooperation between man and machine.

\section*{ACKNOWLEDGMENTS}
This paper is an outcome of discussions of members of the Research Training Group SocialCars
(227198829/GRK1931), funded by the German Research Foundation (DFG).

\bibliographystyle{ieeetr}
\bibliography{mybib}

\end{document}